\documentclass{jetpl}
\twocolumn
\title{Flat band in the core of topological defects: bulk-vortex correspondence in topological superfluids
with Fermi points}

\rtitle{Flat band in the core of topological defects 
}

\sodtitle{Flat band in the core of topological defects: bulk-vortex correspondence in topological superfluids
with Fermi points}

\author{G.E.Volovik
\/\thanks{e-mail: volovik@boojum.hut.fi}
}

\rauthor{G.E.Volovik}

\sodauthor{Volovik}

\address{
 Low Temperature Laboratory, Aalto University, School of Science and
Technology, P.O. Box 15100, FI-00076 AALTO, Finland
\\
Landau Institute for Theoretical Physics RAS, Kosygina 2,
119334 Moscow, Russia
  }

\dates{November, 2010; preprint arXiv:1011.4665}{*}

\PACS{}

\abstract{We discuss the dispersionless spectrum with zero energy in the linear topological defects -- vortices. The flat band emerges inside the vortex  living in the bulk  medium containing topologically stable Fermi points in momentum space. The boundaries of the flat band in the vortex are determined by projections of the Fermi points in bulk to the vortex axis. This bulk-vortex correspondence for flat band is similar to the bulk-surface correspondence discussed earlier in the media with topologically protected lines of zeroes. In the latter case the flat band emerges on the surface of the system, and its boundary is determined by projection of the bulk nodal line on the surface.
 }

\begin{document}

\maketitle

\newcommand  {\version}{v15}
\section{Introduction}

When the fermion zero modes localized on the surface or on the topological defects are studied in topological media, the investigation is mainly concentrated on the fully gapped  topological media, such as topological insulators and superfluids/superconductors of the $^3$He-B type 
\cite{TeoKane2010,SilaevVolovik2010,FukuiFujiwara2010}. However,  the gapless topological media may also have fermion zero modes with  interesting properties, in particular they may have the dispersionless branch of spectrum with zero energy -- the flat band \cite{SchnyderRyu2010,HeikkilaVolovik2010b}.

The  dispersionless bands, where the energy vanishes in a finite region of the momentum space, have been discussed in different systems. Originally the flat band has been discussed in the  fermionic condensate -- the Khodel state 
\cite{Khodel1990,NewClass,Volovik2007,Shaginyan2010}, and for fermion zero modes localized in the core of vortices in superfluid $^3$He-A  \cite{KopninSalomaa1991,Volovik1994,MisirpashaevVolovik1995}. The flat band has also been discussed on the surface of the multi-layered graphene (see \cite{Guinea2006,CastroNeto2009} and references therein). In particle physics, the Fermi band (called the Fermi ball) appears  in  a 2+1 dimensional nonrelativistic quantum field theory which is dual to a gravitational theory in the anti-de Sitter  background with a charged black hole
\cite{Sung-SikLee2009}.  

Recently it was realized that the flat band can be topologically protected in gapless topological matter. It appears in the 3D systems which contain the nodal lines in the form of closed loops  \cite{SchnyderRyu2010} or in the form of spirals \cite{HeikkilaVolovik2010b}. In these systems the surface flat band emerges on the surface of topological matter. The boundary of  the surface flat band  is bounded by the projection of the nodal loop or nodal spiral onto the corresponding surface.  Here we extend  this bulk-surface correspondence to the bulk-vortex correspondence, which relates the flat band of fermion zero modes in the vortex core to the topology of the point nodes (Dirac or Fermi points) in the bulk 3D  topological superfluids.  

\section{Vortex-disgyration}

 As generic example we consider topological defect in 3D spinless chiral superfluid/superconductor of the $^3$He-A type, which contains two Fermi points (Dirac points). Fermions in this chiral superfluid 
 are described by Hamiltonian
  \begin{equation}
H=\tau_3\epsilon(p) +c\left( \tau_1  {\bf p}\cdot {\bf e}_1
+\tau_2  {\bf p}\cdot {\bf e}_2\right)~~,~~\epsilon(p)=\frac{p^2-p_F^2}{2m}
\,,
\label{Hamiltonian}
\end{equation}
where $\tau_{1,2,3}$ are Pauli matrices in the Bogoliubov-Nambu space, and  in bulk liquid the vectors  ${\bf e}_1$ and ${\bf e}_2$ are unit orthogonal vectors.
 There is only one  topologically stable defect in such superfluid/superconductor, since the homotopy group $\pi_1(G/H)=\pi_1(SO_3)=Z_2$.
 We choose the following order parameter in the topologically non-trivial configuration (in cylindrical coordinates ${\bf r}=(\rho,\phi,z)$): 
 \begin{equation}
{\bf e}_1({\bf r})=f_1(\rho)\hat{\boldsymbol{\phi}} ~~,~~ {\bf e}_2({\bf r})=\hat{\bf z}\sin \lambda - f_2(\rho)\hat{\boldsymbol{\rho}}\cos\lambda\,,
\label{OrderParameter}
\end{equation}
with $f_{1,2}(0)=0$, $f_{1,2}(\infty)=1$. 
The unit vector $\hat{\bf l}$, which shows the direction of the Dirac points in momentum space, ${\bf p}_\pm= \pm p_F \hat{\bf l}$, is
 \begin{equation}
~~  \hat{\bf l}({\bf r})=\frac{{\bf e}_1\times {\bf e}_2}{|{\bf e}_1\times {\bf e}_2|}=\frac{f_2(\rho)\hat{\bf z}\cos \lambda + \hat{\boldsymbol{\rho}}\sin\lambda}
 {\sqrt{f_2^2(\rho)\cos^2 \lambda + \sin^2\lambda}}
\,.
\label{l}
\end{equation}

Asymptotically at large distance from the vortex core one has
 \begin{equation}
 \begin{split}
& {\bf e}_1(\rho=\infty)= \hat{\boldsymbol{\phi}} ~~,~~ {\bf e}_2(\rho=\infty)=\hat{\bf z}\sin \lambda - \hat{\boldsymbol{\rho}}\cos\lambda \,,
\\
& \hat{\bf l}(\rho=\infty)= \hat{\bf z}\cos \lambda + \hat{\boldsymbol{\rho}}\sin\lambda\,,
\label{OrderParameterAsymptote}
\end{split}
\end{equation}
which means that changing the parameter $\lambda$ one makes the continuous deformation of the pure phase vortex at $\lambda=0$ to the disgyration in the $\hat{\bf l}$ vector without vorticity at $\lambda=\pi/2$, and then to the pure vortex with opposite circulation at $\lambda=\pi$ (circulation of the superfluid velocity around the vortex core is $\oint d{\bf s}\cdot{\bf v}_{\rm s}=\kappa \cos\lambda$, where $\kappa =\pi\hbar/m$). We consider how the flat band in the core of the defect evolves when this parameter $\lambda$ changes.
 In bulk, i.e. far from the vortex core, the Dirac points are at
 \begin{equation}
 {\bf p}_\pm= \pm p_F \hat{\bf l}(\rho=\infty)  =\pm p_F\left(\hat{\bf z}\cos \lambda + \hat{\boldsymbol{\rho}}\sin\lambda\right)
\,.
\label{DiracPointsAsymptote}
\end{equation}
Due to the bulk-vortex correspondence, which we shall discuss in the next section, the projection of these two points on the vortex axis gives the boundary of the flat band in the core of the topological defect:
 \begin{equation}
E(p_z)=0~~,~~ p_z^2<p_F^2\cos^2 \lambda
\,.
\label{FlatBand}
\end{equation}
This is the central result of the paper: in general the boundaries of the flat band in the core of the linear  topological defect (a vortex) are determined by the projections on the vortex axis of the topologically protected point nodes in bulk. In the next section we consider the topological origin of the flat band and geometrical derivation of its boundaries. In Sec. \ref{QC} ,
the boundaries of the flat band \eqref{FlatBand} are obtained analytically.

\section{Bulk-vortex correspondence}
\label{Bulk-vortexCorrespondence}

Let us first give the topological arguments, which support the existence of the flat band inside the vortex-disgyration line. Let us consider the Hamiltonian \eqref{Hamiltonian} in bulk (i.e. far from the vortex core) treating the projection $p_z$ as parameter of the 2D system.
At each $p_z$ except for two values $p_z=\pm p_F\cos \lambda$ corresponding to two Fermi points (see Fig. 1), the Hamiltonian has fully gapped spectrum and thus describes the effective 2D insulator. One can check that this 2D insulator is topological for $|p_z| < p_F|\cos \lambda|$ and is topologically trivial for $|p_z| > p_F|\cos \lambda|$. For that one considers the following invariant describing the 2D topological insulators or fully gapped 2D supefluids \cite{Volovik2003}:
\begin{equation}
\begin{split}
& \tilde N_3(p_z) 
\\
& =\frac{1}{4\pi^2}~
{\bf tr}\left[  \int    dp_xdp_yd\omega
~G\partial_{p_x} G^{-1}
G\partial_{p_y} G^{-1}G\partial_{\omega}  G^{-1}\right]\,,
\end{split}
\label{N2+1}
\end{equation} 
where $G$ is the Green's function matrix, which for noninteracting case has the form  $G^{-1}=\i \omega - H$. This invariant, which is applicable both to interacting and non-interacting systems, gives 
\begin{eqnarray}
 \tilde N_3(p_z)=1~~,~~|p_z| < p_F|\cos \lambda|
 \,,
\label{TopInsulator}
\\
\tilde N_3(p_z)=0~~,~~|p_z| > p_F|\cos \lambda|
 \,.
\label{Non-TopInsulator}
\end{eqnarray} 

\begin{figure}[h]
\centering
\includegraphics[width=8cm]{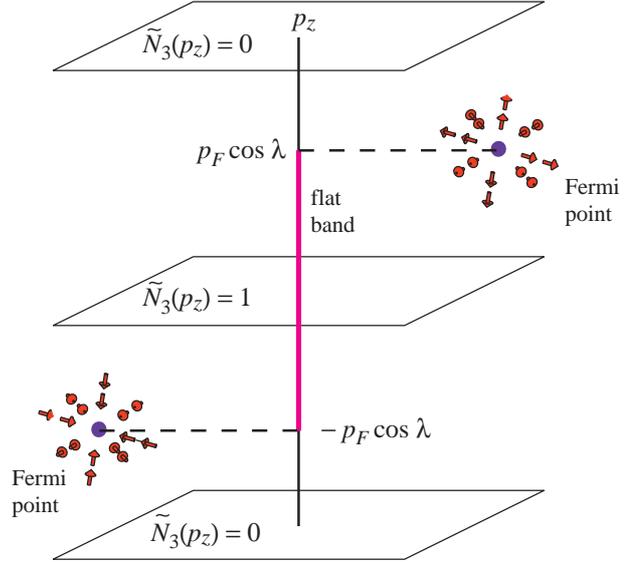}
\caption{
Fig. 1.  Projections of Dirac (Fermi) points on the direction of the vortex axis (the $z$-axis) determine the boundaries of the flat band in the vortex core. Fermi point in 3D systems represents the hedgehog (monopole) in momentum space
\cite{Volovik2003}. For each plane $p_z={\rm const}$ one has the effective 
2D system with the fully gapped energy spectrum $E_{p_z}(p_x,p_y)$, except for the planes
with $p_{z\pm}=\pm p_F \cos\lambda$, where the energy $E_{p_z}(p_x,p_y)$ has a node due to the presence of the hedgehogs in these planes. Topological invariant $\tilde N_3(p_z)$ in 
\eqref{N2+1} is non-zero for $|p_z| < p_F |\cos\lambda|$, which means that for any value of the parameter  $p_z$  in this interval the system behaves as a 2D topological insulator or  2D fully gapped topological superfluid. Point vortex in such 2D superfluids has fermionic state with exactly zero energy. For the vortex line in the original 3D system with Fermi points this corresponds to the dispersionless spectrum of fermion zero modes in the whole interval $|p_z| < p_F |\cos\lambda|$
(thick line).
}
\label{fig:vortex}
\end{figure}

At $p_z  =\pm  p_F|\cos \lambda|$, there is the topological quantum phase transition between the topological 2D ``insulator'' and the non-topological one. The difference of 2D topological charges on two sides of the transition,  $\tilde N_3(p_z=p_F\cos\lambda+0)-\tilde N_3(p_z=p_F\cos\lambda -0)=N_3$,  represents the topological charge of the Dirac point in the 3D system -- hedgehog in momentum space \cite{Volovik2003}.
As we know, the topological quantum phase transitions are accompanied by reconstruction  of the spectrum of fermions bound to the topological defect: fermion zero modes appear or disappear after topological transition in bulk \cite{SilaevVolovik2010,Nishida2010,Nishida2010b,Lutchyn2010}. For the pure vortex, i.e. at $\lambda=0$ or $\lambda=\pi$,  we know from  \cite{KopninSalomaa1991} that   the vortex contains the fermionic level with exactly zero energy for any $p_z$ in the region $|p_z| < p_F$, i.e. in the region of parameters  where the 2D medium has non-trivial topological charge, $ \tilde N_3=1$.  On the other hand no such levels are present after the topological transition to the  state of matter with  $\tilde N_3=0$. 

The similar reconstruction of the spectrum at the topological quantum phase transition takes place for any parameter $\lambda \neq \pi/2$ of the considered defect. This can be understood using the topology in the mixed real and momentum space \cite{GrinevichVolovik1988,SalomaaVolovik1988}. To study fermions with zero energy (Majorana fermions) in the core of a point vortex in a 2D topological superconductor, the Pontryagin invariant for mixed space has been exploited in Ref. \cite{TeoKane2010}.  The Pontryagin invariant describes classes of  mappings $S^2\times S^1\rightarrow S^2$. Here  the mixed space $S^2\times S^1$ is the space $(p_x,p_y,\phi)$, where  $\phi$ is the coordinate around the vortex-disgyration far from the vortex core. This space is mapped  to  the sphere $S^2$ of unit  vector $\hat{\bf g}(p_x,p_y,\phi)={\bf g}(p_x,p_y,\phi)/|{\bf g}(p_x,p_y,\phi)|$ describing the 2D Hamiltonian. In our case it is the  Hamiltonian \eqref{Hamiltonian} outside the vortex core:
 \begin{eqnarray}
 H_{p_z,\lambda}(p_x,p_y,\phi)=\tau_i g^i(p_x,p_y,\phi;p_z,\lambda)\,,
 \label{Hamiltonian_g}
\\
 g^3=\frac{p_x^2+p_y^2}{2m}-\mu(p_z)~~,~~\mu(p_z)=\frac{p_F^2-p_z^2}{2m}\,,
\nonumber
\\
 g^1=c(p_y\cos\phi -p_x\sin\phi) \,,
 \nonumber
\\
 g^2=c(p_z\sin\lambda -\cos\lambda(p_x\cos\phi +p_y\sin\phi))
\,,
\label{g-vector}
\end{eqnarray}
with $p_z$ and $\lambda$ being the  parameters of this effective 2D Hamiltonian.
The Pontryagin $Z_2$ invariant is non-trivial and thus the zero energy state exists in the  core of the defect the effective 2D superconductor, if  the parameters $p_z$ and $\lambda$  of the 2D Hamiltonian \eqref{Hamiltonian_g} satisfy condition $|p_z| < p_F|\cos \lambda|$.

For the considered linear topological defect (vortex-disgyration) in the 3D system this implies that the core of this defect contains the dispersionless band in the interval  of momentum $|p_z| < p_F|\cos \lambda|$, i.e.  one obtains equation \eqref{FlatBand}.

\section{Flat band from quasi-classical approach}
\label{QC}

Let us now support the above topological arguments by explicit calculation of the fermionic flat band in the vortex-disgyration, which is described by the order parameter \eqref{OrderParameter}. The  Bogoliubov-de Gennes Hamiltonian for fermions localized on the defect line is obtained from \eqref{Hamiltonian} by substitution of the classical transverse momentum by the  quantum-mechanical operator, 
 \begin{equation}
 {\bf p}_\perp \rightarrow  (-i\nabla_x, -i\nabla_y)  \,,
\label{MomentumOperator}
\end{equation}
while $p_z$ remains the good quantum umber which serves as parameter of the effective 2D system. The zero energy states in this 2D system can be studied using the quasiclassical approximation, see details in Chapter 23 of the book \cite{Volovik2003}. For our purposes it is sufficient to consider the Hamiltonian on the trajectory $s$ which crosses the center of the vortex. The modification of quasiclassical Hamiltonian in Eq.(23.16) in \cite{Volovik2003} for the considered vortex-disgyration is
 \begin{equation}
 \begin{split}
& H_{\rm qcl}(p_z)=-i\frac{q}{m}\tau_3\partial_s +U(s) \tau_2 \,,
\\
& U(s)= cp_z\sin\lambda - cq f_2(|s|) {\rm sign} (s) \cos\lambda
\,,
\\
& q=\sqrt{p_F^2 -p_z^2} \,.
\end{split}
\label{QuasiclassicalH}
\end{equation}
The Hamiltonian $H_{\rm qcl}(p_z)$ is super-symmetric if the asymptotes of the potential $U(s)$ have different sign for $s=- \infty$ and $s=+ \infty$. The latter takes place if 
 \begin{equation}
|p_z|\sin\lambda < q|\cos\lambda| \,.
\label{Constraint}
\end{equation}
The super-symmetric Hamiltonian $H_{\rm qcl}(p_z)$ has the state with zero energy, $E_{\rm qcl}(p_z)=0$, for any $p_z$ in the interval \eqref{Constraint}. For vortices in chiral superfluids it is known \cite{Volovik2003} that the zero energy state of the quasiclassical Hamiltonian, $E_{\rm qcl}=0$, automatically results in the true zero energy state, $E=0$,  obtained  in the exact quantum-mechanical problem using the Bogoliubov-de Gennes Hamiltonian. This proves the existence of the flat band in the range of momentum \eqref{Constraint}, which coincides with equation \eqref{FlatBand} and is in agreement with the topological analysis in previous section. 

\section{Discussion}

We discussed the 3D matter with  topologically protected Fermi points. Topological defects (vortices and vortex disgyrations) in such matter contain the dispersionless fermionic band with zero energy -- the flat band.  The boundaries of the flat band are determined by projections of the Fermi points on the axis of the topological defect. This bulk-vortex correspondence for flat band is similar to the bulk-surface correspondence discussed in the media with topologically protected lines of zeroes \cite{HeikkilaVolovik2010b,SchnyderRyu2010}. In the latter case the  flat band is formed on the surface of the system, and its boundary is determined by projection of the nodal line (closed loop \cite{SchnyderRyu2010} or spiral \cite{HeikkilaVolovik2010b}) on the corresponding surface.

This work is supported in part by the Academy of Finland, Centers of excellence program 2006--2011.
It is my pleasure to thank V.B. Eltsov, T.T. Heikkil\"a and N.B. Kopnin for helpful discussions.

\end{document}